\documentclass[12pt]{article}
\usepackage[left=2cm,top=2cm,right=2cm,bottom=2cm,footskip=15pt,nohead]{geometry}
\usepackage{graphicx}
\usepackage{hyperref}
\usepackage{setspace}
\usepackage[numbers]{natbib}   
\usepackage{amssymb}
\usepackage{verbatim}
\usepackage[usenames,dvipsnames]{color}
\hypersetup{colorlinks=true, citecolor=black, linkcolor=black, citecolor=black, urlcolor=black, filecolor=black}

\newcommand{\commentout}[1]{}

\newcommand{\baseloc}{../../../}

\begin{document}

\title{The Social Contagion Hypothesis: Comment on ``Social Contagion Theory: Examining Dynamic Social Networks and Human Behavior''}
\author{A.C. Thomas}
\date{July 9, 2012}
\maketitle

\onehalfspacing

\begin{abstract}
I reflect on the statistical methods of the Christakis-Fowler studies on network-based contagion of traits by checking the sensitivity of these kinds of results to various alternate specifications and generative mechanisms. Despite the honest efforts of all involved, I remain pessimistic about establishing whether binary health outcomes or product adoptions are contagious if the evidence comes from simultaneously observed data.
\end{abstract}

\section{Introduction: Correlation and Influence}

The works of Profs. Christakis and Fowler (henceforth ``CF'') have provoked a great deal of discussion about how the social network of a society might impact the health outcomes of its members, and they should be applauded for bringing these social hypotheses into public discussion, as well as for making the social network aspects of the Framingham Heart Study available for investigation. 

In doing so, they remind us that while causal inference and network science are both difficult areas with their own rich methodological histories, causal inference on network processes provides a whole new level of difficulty. The conclusions CF raise are indeed extraordinary and demand an equally thorough analysis to check that all assumptions have been covered. Here I seek to clarify those areas in which the methods are adequate or deficient.

In each of their observational studies, the general structure of their arguments has three steps:

\begin{enumerate}

\item Establish correlation of behaviours for the members of a network, and ensure statistical significance of this correlation through comparison to a null distribution. 

\item Show that some fraction of this autocorrelation is in fact due to a directional mechanism in the network, even in the presence of confounding factors and processes.

\item Hypothesize an actual causal mechanism at work for future investigation.

\end{enumerate}

The strongest elements of these studies are invariably in Step 1, and with a completely specified network, these correlational assessments are essentially beyond reproach in establishing that network effects are present. Step 3, hypotheses for why a pattern might be observed in a particular way, is editorial in nature and is reserved either for experimentation or finer-scale observation to be tested. It is Step 2 where many of these difficulties lie, particularly with regard to the sensitivity of the results to model specification and missing data.

Each of these gives rise to the following points, which I demonstrate here:

\begin{enumerate}

\item The word ``influence'' should be used exclusively in the causal manner in which it has been historically known.

\item The net temporal peer effect of many of these studies is indistinguishable from zero, even if some components are estimated to be positive and significant, so that most observed correlations are contemporary and not time-ordered.

\item Contemporaneous models for causation introduce more problems than they solve in the social setting.

\item Most dramatically, as implemented, the asymmetric contemporaneous causation test for binary outcomes has considerable unresolved theoretical problems.

\end{enumerate}

\section{Associational Interpretations of Network Autocorrelation}

Network autocorrelation has many different names -- clustering, association and peer effects are among those used in sociology and economics -- but effectively encompasses the sum of all network processes that we seek to identify. This includes, but is not limited to, manifest and latent homophily (or heterophily), external common factors, and social contagion, also well-known in this context as social influence or induction. 

While each of these pieces has an explicit mechanism of causation (see \citet{shalizi2011hacagcosns} for a prior discussion), it is the possibility that the lion's share of this autocorrelation is due to contagious spread, yielding the concept of ``three degrees of influence''. This provides a common hypothesis to much of the joint work of CF. The analogy of a ``ripple effect'', made by both Christakis and Fowler in public presentations, seems much more preferable in hindsight as the preferred popularized explanation of why network effects are so important. \footnote{My first exposure to this idea was in the public media: the 1993 series finale of the television show Quantum Leap, in which one man's hard work and influence, carefully applied and targeted, has in fact caused ripples through the people of the world that changed it for the better. I have no doubt that CF had similar cultural influences to their ideas.}

Here I fear the (literally) hypothetical use of the word ``influence'' has manifested its own undue effect. For example, in their monograph \citep{christakis2009csposnahtsol}, CF immediately jump to the term ``three degrees of influence'' when describing the extent of network autocorrelation, whether or not we may also call it clustering, association, correlation or peer effects. while I understand that a choice of wording is ``evocative'', this basis is needlessly confusing. The word ``influence'' is a well-established proxy for a directional causal effect in every scientific and sociological context in which it appears (including every citation in the CF discussion paper containing the word ``influence'' in the title.)

There is also a lingering question: is this ``three degrees of autocorrelation'' (admittedly, far less snappy) the result of a process that actually acts across three sequential network connections? Because this marginal measure of correlation does not explicitly control for intermediaries, it is possible (if not outright likely) that the observed degree of correlation is more than that of the individual effects that produce it. In their experimental works, CF do indeed provide compelling evidence that the autocorrelation of a trait is caused by contagion, as well as its degree to which the effect ripples, but this is due not only to an extremely plausible mechanism (such as the change of behaviour during a game), but also to randomized assignment and to a time-ordered sequence of events. 

Indeed, this distinction is nearly impossible to make in the static case. If Irene has two friends Jack and Kevin, who are not friends with each other, then {\it any} mechanism that produces correlation between Irene and each of her friends will also produce it between Jack and Kevin, no matter what direction the arrows point; two degrees of autocorrelation is manifested from a process that only takes one step. As CF have themselves pointed out, being able to distinguish multiple ripples of contagion in an observational task is already difficult without this complication.

As an exploratory means of developing hypotheses, I agree that the actual ripple count is secondary to establishing the existence of the effect itself. Indeed, no study based on the Framingham Heart Study has attempted to explicitly prove causative contagion beyond one degree. The challenge remains whether or not these methods are sufficient to show causation for a single step.

\section{Simultaneity of Events in the Network Setting}

The basic form of the CF estimation procedure is a logistic regression for an ego's outcome at each time point. If $Y_{i,t}$ is the outcome for person $i$ at time $t$, and $W_{ij}$ is the network term where person $i$ names person $j$, the general equation form preferred by CF is 

\begin{equation}\label{eq.1}
\log \frac{P(Y_{i,t+1}=1)}{P(Y_{i,t+1}=0)} = \mu + \alpha Y_{i,t} + \beta \sum_j W_{ij,t} Y_{j,t} + \gamma \sum_j W_{ij,t+1} Y_{j,t+1} + \delta X_{i,t+1} 
\end{equation}
so that $(\mu, \alpha, \beta, \gamma)$ represent the baseline, autocorrelative, time-lagged peer and time-concurrent peer coefficients respectively, and $(\delta, X_{i,t+1})$ are coefficients and exogenous covariates. 

CF's analyses appears to take a slightly different form in two ways, with both involving the consideration of one dyad at a time. First, they model only one type of tie at a time, rather than including both directions of tie explicitly in the same formula (unless the tie is one of mutual friendship). Second, they consider each potential dyad as its own outcome -- that is, each outcome for an individual is repeated by the number of ties in which they participate -- and use Generalized Estimating Equations to deal with this correlation in observations. This is clearly not a generative probability model, but given the relative sparseness of the FHS friendship network, the distinction may be minor. (See Section \ref{s:simulation-study} for more on why this is the case.)

CF present the time-concurrent peer coefficient as evidence of the contagion effect, and the time-lagged peer coefficient exclusively as a control for manifest homophily. This restriction in model choice -- defining the causal effect as it pertains to contemporaneous variables exclusively -- is what I feel to be the most debatable point in all of CF's analyses, and one that deserves the highest level of scrutiny.

\subsection{Product Adoption As Social Contagion}

The first consideration should be the source of the theory that justifies the choice model. CF reference \citet{valente2005nmamfsdi}, who deals with a comparatively simple problem: whether a person adopts a product for use (in this case, contraceptive use). And that this adoption is one way only; a person classified as a user remains a user from this point onward. This means that the formula

\[ \log \frac{P(Y_{i,t+1}=1|Y_{i,t}=0)}{P(Y_{i,t+1}=0|Y_{i,t}=0)} = \mu + \beta \sum_j W_{ij,t} Y_{j,t} + \gamma \sum_j W_{ij,t+1} Y_{j,t+1} + \delta X_{i,t+1} \]
can only include the outcomes for individuals who had {\em not} adopted at time $t$ (since the probability of adoption, given that one has already adopted, is one), though this formula also explicitly considers the change in a reported network. Quoting this article (with notation changed to match),

\begin{quote}
a positive and significant $\beta$ indicates that respondents with high network exposure at baseline were more likely to adopt at time $t+1$. A positive and significant $\gamma$ indicates that change in network exposure is associated with change in behavior. This may indicate contagion, but still may be a product of some omitted factor.
\end{quote}

It is clear that both terms $\beta$ and $\gamma$ indicate that network exposure affects adoption, which is is the textbook definition of contagion. Moreso, the term from the past is {\em unambiguously} so, if there is no other confounding. In fact, the statement on the latter is incomplete: by its definition, $\gamma$ refers to a change in exposure if the network term $W_{ij}$ and/or peer term $Y_j$ changes between $t$ and $t+1$. In each case, the implication is clear: total exposure, not average exposure among friends, is the contagious factor at work in these methods. It may of course be that each network tie $W_{ij,t}$ has a value proportional to the actual strength of a friendship (if not downright preferred -- see \citet{thomas2010vttfl} for a more thorough discussion). The notion that two mutual directional ties should be stronger than a single directional tie is evidence of this alone; constraining the total social autocorrelation to be the same for each individual is a design choice that is uncommon in the product adoption literature.

While it may be possible to find evidence involving contagion in only one of these terms, matters get tricky when we disregard the other. The persistence of the state of obesity in the data is strong; roughly 90\% of individuals stay thin or fat between examinations, so that the lagged predictor will be highly correlated with the contemporaneous one. The differences between exposure and change in exposure may be difficult enough to distinguish without this complication.

\subsection{The Sum of Peer Effects Over Two Time Points is Effectively Zero}\label{s:netzero}

In its purest form, with all assumptions satisfied, the regression equation gives a counterfactual interpretation of how an outcome will change if one of its predictors were substituted with another value. The total effect of exposure to peer obesity, for example, would be the sum of all coefficients for obesity at all included time states, as this would be the equivalent exposure to a peer with that trait for the entire time. 

There is a consistent pattern of the peer coefficients in many of CF's results: the contemporaneous coefficient is positive and significant, while the lag-one coefficient is negative, significant and roughly the same size as its counterpart, so that they sum to nearly zero. If the prior lag term was only controlling for homophily, we would expect this effect to be positive; in very few cases, this is so. Table \ref{t:effect-summaries} gives four examples for the total correlation of a mutual friend's state; only loneliness has effects of the same sign.

\begin{table}
\begin{center}
\begin{tabular}{cccc}
\hline
\hline
Behavior (Reference) & Lag-1 Ego & Contemporary Alter & Lag-1 Alter \\
\hline
Obesity (Table S1) & 4.35 & 1.19 & -1.25 \\
Smoking (Table S8) & 4.49 & 0.51 & -0.53 \\
Happiness (Table S6a) & 3.19 & 2.07 & -1.87 \\
Loneliness (Table 5a) & 0.28 & {\bf 0.41} & {\bf 0.16} \\
\hline
\hline
\end{tabular}
\caption{\small The coefficient estimates for ``mutual friends'' in several CF studies. Three of the four are highly anti-correlated pairs that nearly cancel out, suggesting that the sum of peer influence in the evolution of these trends is negligible. These states are also quite persistent over time in the ego, as shown by their own lag. \label{t:effect-summaries}}
\end{center}
\end{table}

What does this mean for many of these studies, at a minimum for obesity, smoking and happiness? It speaks to a very specific type of change in network behaviour: there is {\bf only} a nearly-simultaneous effect at play here; $Y_{i,t+1}$ and $Y_{j,t+1}$ change together from zero to one, or one to zero, in terms of the observed effect. \citet{cohencole2008dis} point out that this type of effect -- nearly equal-sized effects with opposite magnitude -- can be produced by collinearily and confounding in the predictors. In the case of obesity, smoking and happiness, this is clearly the case; the lag-1 and contemporary predictors for the alter's state are highly correlated. This means that the weight of the difference in the regression is being produced by the slight differences between these states -- that is, for a change in peer behaviour, not for overall exposure.

This can be mitigated by choosing predictors that are minimally collinear, or better yet, correspond to the process believed to be in place -- that {\em changes} in these states are autocorrelated on the network. If there is any sort of a causal peer effect taking place, it must be at a time scale well below the frequency of observation (and, short of running a new observational study, well beyond our control). Still, the story is clear: if these traits were clearly ``spreading'' (or receding) on social networks, then baseline exposure would seem to be a necessary predicate.

\section{Observed Correlation Is Sensitive To Dichotomization Choice}

Obesity itself is not a strictly dichotomous trait, but a clinical definition with respect to the continuous body mass index. As a binary trait, it becomes an easier tool for clinicians to use in diagnosis. It also makes the adoption mechanism much clearer to model: either someone has the trait or they don't, and its ``spread'' is far easier to quickly tabulate. (Any form discretization will accomplish the same goal, though the mechanisms at work will be more complicated to understand.) 

\begin{figure}
\begin{center}
\includegraphics[width=0.8\linewidth]{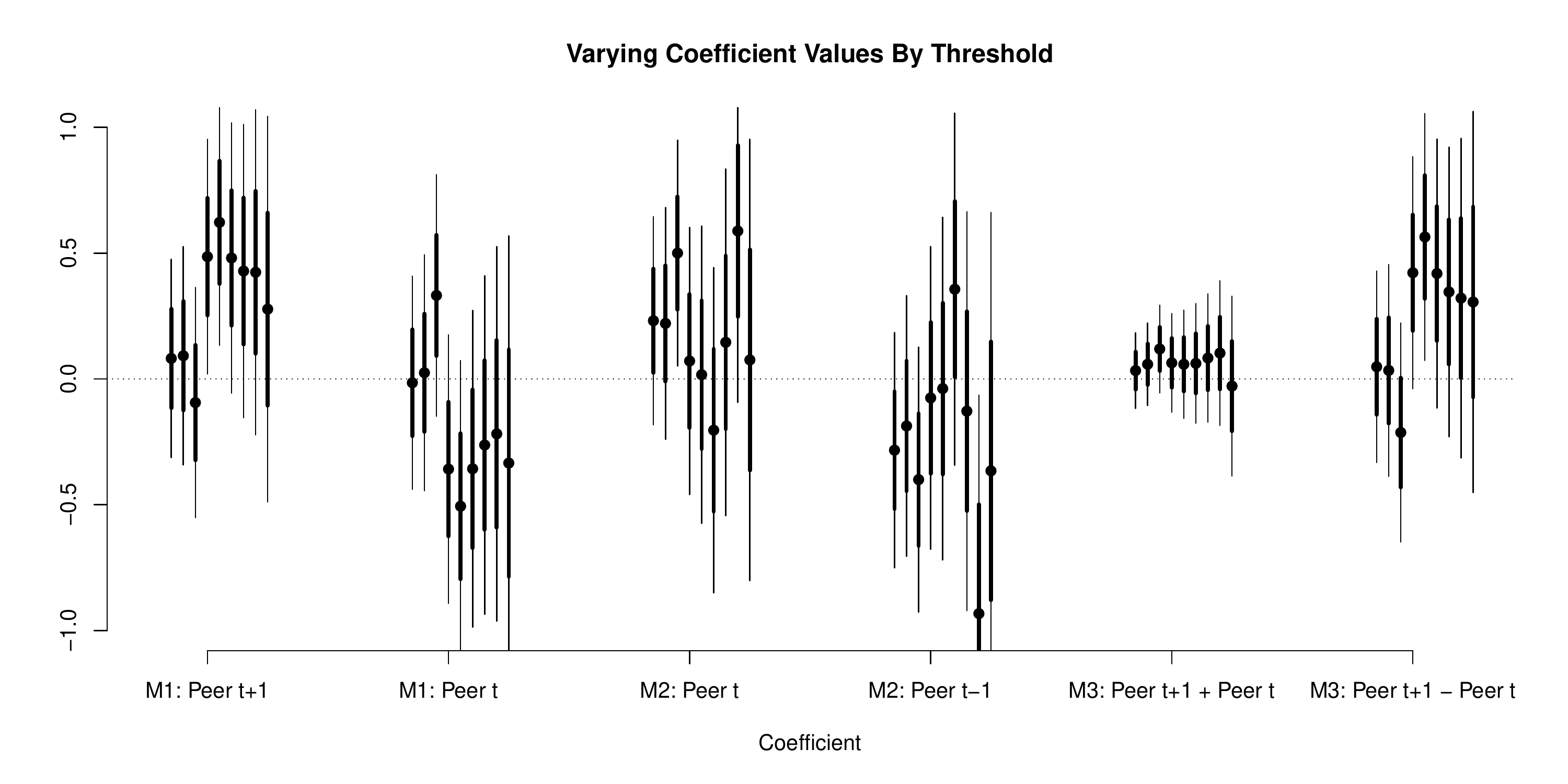}
\end{center}
\caption{\small Various coefficient estimates taken on three models of the FHS Network Obesity Study (M1, M2 and M3). Models 1 and 2 have two alter lags; Model 3 has the sum and difference of Model 1's predictor lags. The sum is considerably more stable to perturbation of the dichotomizing threshold. The difference, which corresponds to the change in network exposure, is more sensitive to the choice of cut point. \label{f:variability}}
\end{figure}

To check the validity of the assumption, I run three separate analysis with named ties: the standard CF set-up with alter contemporaneous and lag 1, the modified set-up with alter at lag 1 and 2, and a third setup where the predictors are the sum and difference of the alter contemporaneous and lag 1 states. (All three analysis had one ego lag included as well.) For each analysis, I dichotomized the condition for ``obese'' differently, from a threshold BMI of 28 to 32 every 0.5 units. Figure \ref{f:variability} shows the estimated coefficients in each case, and it is clear that there is considerable change in the predicted coefficients across this range. Moreover, in the first two pairs, the complementary coefficients can be seen to be negatively correlated as in the original case.

For the third analysis, the predictor for the sum of the obesity states is nearly zero in each case. It is in the difference where this shows up; that is, it is not exposure to obesity that is correlated with ego obesity, but a change in alter obesity that shows any positive pattern -- and even there, the effect is maximized at a cutoff of 30.

\subsubsection*{Do Thin Friends Make You Thin? Effect Directionality in Modelling Choice}

Any dieter will tell you that the process of losing weight is far different than the process of gaining it. The product adoption model gives a nice benefit that the standard CF logistical model does not: one can treat adoption and rejection as two separate processes, since the data are already stratified on this basis. The rejection model is simply taken with units that had adopted at time $t$:

\[ \log \frac{P(Y_{i,t+1}=1|Y_{i,t}=1)}{P(Y_{i,t+1}=0|Y_{i,t}=1)} = \mu + \beta \sum_j W_{ij,t} Y_{j,t} + \gamma \sum_j W_{ij,t+1} Y_{j,t+1} + \delta X_{i,t+1} \]

This is of course a subset of the original model with directional predictors added, interacted with the ego's prior state, but the story it tells is clearly different and worth mentioning.

I ran each model from the obesity paper using the replication data provided through the SHARE database. The peer effect estimates do not change substantially for the adoption of obesity, but new results for the loss of obesity show similar autocorrelative behavior and a higher rate of retention in the obese state.

\section{Finding Causal Stories in Simultaneous Data}

The argument from asymmetry is one of the most compelling notions of the CF analyses. By this argument, for each friendship named, there are two possible effects: the ``causal'' direction, where the namer is thought to take influence from the named individual, and the ``noise'' direction, which goes opposite. If the difference is positive and significant, then this is evidence of a contagious effect. If there are non-contagious autocorrelations present, then both of these effects should be estimated to be positive. It is clear that the coefficient order ``mutual $>$ named $>$ namer'' is a consequence of this simpler case, and that the mutual tie should have a stronger autocorrelation than either named or namer is due to the sum of these autocorrelations as an effectively stronger tie.

This argument is by far at its most robust if it can be attached to a causal model with a coherent, generative probability distribution \citep{lyons2011semvfsa}. This is one reason why \citet{shalizi2011hacagcosns} avoids the problem by dealing with predictors that must have come before the observed period (and which are consistent with the characteristic time scale of the process under study). For this method to be valid on contemporaneous data, there are at least two seemingly viable options:

\begin{enumerate}


\item Use the quasi-experimental methods proposed by \citet{rubin1974ecetrans} to eliminate bias due to imbalance, for which we can distinguish latent homophily from contagion by matching experimental units on all visible characteristics (see \citet{aral2009distinguishing} for one such study). However, this still requires the knowledge and plausible timing of the ``treatment'', in this case the exposure to the behavior or trait; it merely reduces one sort of uncertainty by using observed variables in a new way. Even using this matching procedure will not deal with any unmeasured factors.

\item Use the causal structural modeling methods of \citet{spirtes2001cpas} and \citet{pearl1995cdfer}. This would by far have the most promise if we had a way to embed this process in a directed acyclic graph. This would be ideal if we had pairs of friends only and randomized (or held ignorable) assignment of declared friendship traits. 

\end{enumerate}

\subsection{Standard Ising Models Are Unidentifiable For Direction}
One method for producing simultaneous outcomes on a network is a static Ising model; the joint probability distribution over all nodes is proportional to the total energy in the system. Consider a binary outcome where any individual $i$ can take a value $Y_i \in \{0, 1\}$. In one simple form, the energy level for one node is calculated from three parameters:

\[ E(i) = \alpha Y_i + \beta \sum_{j \neq i} W_{ij}(Y_i - Y_j)^2 + \gamma \sum_{j \neq i} W_{ij} Y_iY_j. \]

A positive $\alpha$ gives higher energy to an individual with a positive value of the trait. A positive $\beta$ gives higher energy if individual $i$ and its neighbours have differing values. Finally, a positive $\gamma$ raises the energy if adjacent individuals both have a positive attribute.

The probability distribution over the ensemble is then the Gibbs distribution,

\[ P(\mathbf{Y}|\alpha, \beta, \gamma) = \exp(\sum_i E(i)) / Z (\alpha, \beta, \gamma), \]
where $Z(\alpha, \beta, \gamma)$ is the normalizing constant, calculated by summing over all $2^n$ possible configurations of the network outcomes. Draws from this model can be obtained through Markov Chain Monte Carlo for known values of $(\alpha, \beta, \gamma)$.

If all network ties $W_{ij}$ are also zero or one, then the directional effect is unidentifiable. One can see that the total energy is

\[ E_{total}^1 = \sum_i \alpha Y_i + \beta \sum_i \sum_{j \neq i} W_{ij}(Y_i - Y_j)^2 + \gamma \sum_i \sum_{j \neq i} W_{ij} Y_iY_j, \]
which has the same value if we use the reverse-direction edge 
\[ E_{total}^2 = \sum_i \alpha Y_i + \beta \sum_i \sum_{j \neq i} W_{ji}(Y_i - Y_j)^2 + \gamma \sum_i \sum_{j \neq i} W_{ji} Y_iY_j, \]
since the $\beta$ and $\gamma$ terms are identical in summation, since their respective squared-difference and product terms are themselves symmetric. The Ising model can be used to generate an ensemble of network outcomes, but cannot distinguish between the directions of ties in this case.

\subsection{Simultaneous Autoregressive Models}

We may well have a plausible causal mechanism in the Simultaneous Autoregressive Model (see for example \citet{doreian1982mlmflmseasdt}), beginning with a multivariate Gaussian distribution defined as

\[ Z_i = \rho \sum_j W_{ij} Z_j + \epsilon_i, \]
or in its explicit multivariate form,
\[ \mathbf{Z} = \rho \mathbf{WZ} + \mathbf{U}, \]
where $\mathrm{Cov}(U)$ is a matrix $\Sigma$; solving for $\mathbf{Z}$ yields $\mathbf{Z} = (\mathbf{I}-\rho \mathbf{W})^{-1}\mathbf{U}$. A power series expansion of the leading term gives

\[ \mathbf{Z} = (\mathbf{I} + \rho \mathbf{W} + \rho^2 \mathbf{W}^2 + \rho^3 \mathbf{W}^3 + ...) \mathbf{U}; \]
that is, this model describes some set of initial disturbances $\epsilon_i$, which need not be autocorrelated, that are spread throughout the network, diminishing at each step. Even if the link process is asymmetric, the resulting distribution is itself symmetric by necessity, whether or not contagious processes, manifest or latent homophilies, or external common causes were at work. The asymmetry comes from the fraction of the variance in an individual's outcome that is shared by those they named, which is greater than the share for those that named them. While a time-based mechanism would go much farther to establish the actual causal mechanism, and while things would be simpler without the complication of cycles or mutual ties, this is at least a reasonable starting point: we can obtain a maximum likelihood estimate for $\rho$ rather than a regression estimate. Better yet, we can take the two-way model

\[ Z_i = \rho_1 \sum_j W_{ij} Z_j + \rho_2 \sum_j W_{ji} Z_j + \epsilon_i, \]
where $\rho_1$ and $\rho_2$ characterize adoption from those the ego named and those who name the ego respectively; the difference may then categorize the asymmetric condition appropriately even if its estimates of $\rho_1$ and $\rho_2$ are biased.

The catch is that this model, and its associated proofs and guarantees, come from its construction as a valid, coherent probability model with a causal story -- and deliberately constrict the sorts of conclusions one can draw from it. The specification of covariates alone is tricky -- for example, does a named friend's smoking status affect the ego's obesity level, and how will collinearily and confounding come into play here? Here, at least, there's a way forward if we have exogenous predictors \citep{bramoulle2009ipetsn}.

Rather than using the MLE to calculate each $\rho$, a typical method to estimate the $\rho$ parameters is by simple and direct regression. This is the approach used by CF and is well-known as the QAD (``quick-and-dirty'') method \citep{doreian1984namsmcr}. This is known to produce biased estimates since it is incoherent as a probability model; there is explicit dependence between the rows of the model that is not handled in a principled fashion (even absent the apparent ``off-label'' use of GEEs present in these papers: repeating the same ego outcome at the same time point, given different alters under inspection, seems to deviate wildly from the repeated-measurements-over-time method, the original intent of \citet{liang1986ldauglm}.) Of course, just because a model gives biased estimates is not a sufficient reason to discard it from use; guarantees of consistency from the original model do allow for some leeway. 

We forfeit all these guarantees once we adopt a binary outcome with an incoherent probability model. The properties of the estimators of these models are unknown to theory, including how they behave under different network specifications. Among other properties, under the standard Gaussian SAR model, the marginal variance of each observation depends on its position in the network; degree has by far the biggest impact on this for even modest autocorrelation levels. This heteroscedasticity has the potential to bias estimates of all sorts if not properly handled -- indeed, the fact that ``influenced'' units have greater variance, and a greater share of their variance attributed to their peers, is at the very heart of the asymmetry method.

\subsection{Simulation Study}\label{s:simulation-study}

To demonstrate how the specification of the network can affect these estimates, I conduct a series of simulations on networks specified with the SAR model and estimated with standard unadjusted quick-and-dirty regression methods. The simulation focuses on how network geometry alone affects estimates of autocorrelation and asymmetry, and does not use any homophily in the tie formation process.

The simulation steps are relatively straightforward:

\begin{figure}
\begin{center}
\includegraphics[width=\linewidth]{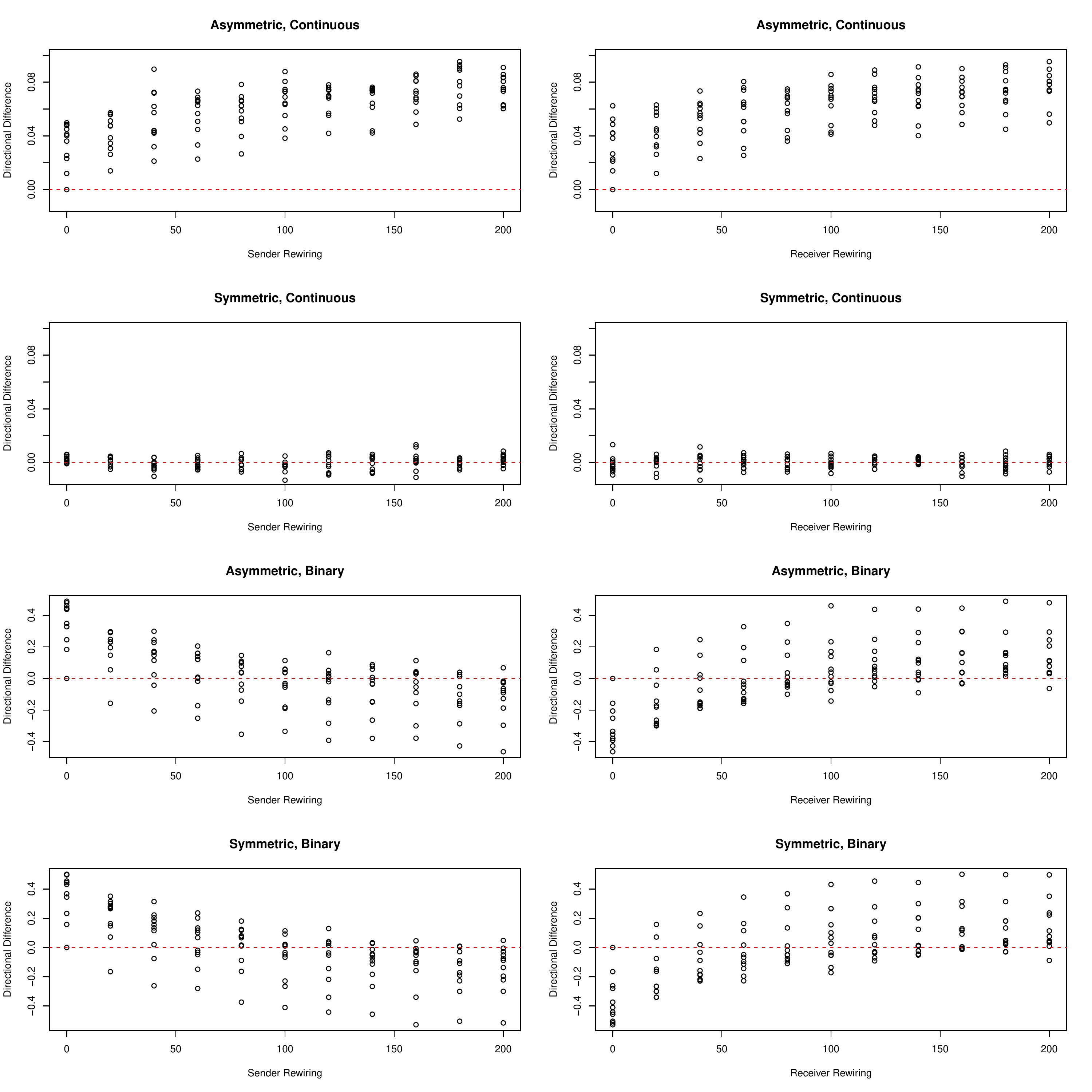}
\end{center}
\caption{\small The asymmetry condition as measured by quick-and-dirty linear (rows 1 and 2) and logistic regression (rows 3 and 4). The test works with linear continuous data and greater heterogeneity in indegree and/or outdegree (panels 1 and 2), and does not detect asymmetry in symmetric data (panels 3 and 4). The test with binary data gives a positive result with heterogeneous sender rewiring, and a negative result with heterogeneous receiver rewiring, whether or not the process is symmetric or asymmetric. \label{f:perturb-net-sims}}
\end{figure}

\begin{itemize}

\item Generate a directed-edge network so that all nodes have the same indegree and outdegree. (For the sake of this analysis I use a 200 node network with outdegree 1, similar to the number of egos with declared friends in the SHARE database.)

\item Rewire a number of edges by changing their receiver, thus changing the distribution of node indegree. (This is done merely to introduce variability in the network structure and need not be connected to homophily.)

\item Rewire a number of edges by changing their sender, thus changing the distribution of node outdegree.

\item Generate two continuous outcomes $\mathbf{Z}^1$ and $\mathbf{Z}^2$ by the SAR model: one with the intended directional matrix alone, one with identical action in each direction. (In case 1, $(\rho_1, \rho_2) = (0.4, 0)$; in case 2, $(\rho_1, \rho_2) = (0.2, 0.2)$.

\item Generate a binary version of each outcome, $\mathbf{Y}^1$ and $\mathbf{Y}^2$,  by dichotomizing the outcome at zero (the true median of this process, though this value is flexible.)

\item Conduct quick-and-dirty linear and logistic regressions with each directional network term (forward and reverse) as predictors.

\item Calculate the asymmetry measure by subtracting the reverse coefficient from the forward coefficient.

\end{itemize}

Replication code is included as a supplement to this document; the results of this are summarized in Figure \ref{f:perturb-net-sims}. Each point represents the mean result of 100 simulated outcomes with one rewired network. The simulations reveal a very interesting difference between the continuous and binary models. The symmetry test works exactly as advertised for the continuous model, where performance is enhanced by additional variability in both indegree and outdegree, and no asymmetric effect is detected in the symmetric model. However, the test completely fails to discriminate between asymmetry and symmetry in the binary case. There is absolutely no distinction between the patterns in the two test cases. The trend is driven entirely by the variabilities in degree, and hence the structure of the network, whether or not the autocorrelation was produced by an asymmetric or symmetric mechanism. This asymmetric pattern is also observed when data are generated from the static Ising model, whose results are symmetric by construction even if the ties are asymmetric.

Moreover, the effect is measured to be positive when there is minimal sender rewiring, and maximal receiver rewiring -- that is, when outdegree is relatively similar and indegree is more variable, or exactly the situation that the ``name one friend'' Framingham mechanism produces naturally. \footnote{Replications of network asymmetries by other authors use networks with similar naming mechanisms. In many of the schools in the Add Health study, students are still asked to name (at most) one or two friends, leading to the same constrained out-degree versus a more variable in-degree.}

\begin{table}
\begin{center}
\begin{tabular}{l|ccccccc}
\hline
P(Difference $>$ 0) & Wave 1 & Wave 2 & Wave 3 & Wave 4 & Wave 5 & Wave 6 & Wave 7 \\
\hline
Asymmetric, continuous & 0.893 & 0.885 & 0.898 & 0.937 & 0.949 & 0.935 & 0.924 \\
Symmetric, continuous & 0.505 & 0.489 & 0.519 & 0.491 & 0.499 & 0.504 & 0.510 \\
Asymmetric, binary & 0.630 & 0.617 & 0.660 & 0.686 & 0.743 & 0.706 & 0.705 \\
Symmetric, binary & 0.639 & 0.646 & 0.650 & 0.715 & 0.727 & 0.727 & 0.685 \\
\hline
\end{tabular}
\caption{\small The results of 1000 simulations using the FHS Social Network with asymmetric and symmetric effects, with continuous and binary outcomes under a SAR generative model. Each cell represents the fraction of instances when the asymmetry comparison is greater than zero. While it is clear that the continuous data yields a valid test (with fractions considerably above 0.5 for asymmetric generation, and equal to 0.5 in the symmetric case), there is virtually no distinction between the binary cases; 4 of 7 waves have higher outcomes for the symmetric case.\label{t:asymm-fraction}}
\end{center}
\end{table}

Of course, simulations alone do not mean that this pattern must be present in real network data. To test that the Framingham data can produce asymmetric results from symmetric models, I repeat this test with the actual SHARE network as declared for friendships in each wave of the study, simulate 300 outcomes by the continuous SAR model for the directed and symmetric cases, and run the quick-and-dirty regressions on these data sets. As shown in Table \ref{t:asymm-fraction}, the same pattern holds; the number of times the asymmetry measure is greater than zero is virtually identical for the binary cases.

\subsection*{Consequences to the CF methodology}

The results of these simulations do not necessarily mean that there is no contagion present in any of these systems. But even though the principles of the CF approach are preserved with this method -- the asymmetry test, on contemporaneous binary data, with a quick-and-dirty regression -- the fact that this method has no power to detect asymmetry is startling. And since the same friendship network is used in each of the CF papers, this raises the unfortunate possibility that many if not all of the positive asymmetry effects obtained in these analyses are due not to causal effects, but to simple autocorrelation through homophily, whose estimates are modulated by an unbalanced network. 

How can we possibly tell the difference? For a start, there are coherent probability models for correlated, contemporaneous binary data; for example, the auto-probit method takes the SAR specification for a continuous $Y$ and sets a binary variable $W_i = \mathbb{I}(Y_i > c)$ for some constant threshold value $c$ (typically zero). However, it is considerably more computationally burdensome to solve for autocorrelation parameters than with a quick-and-dirty regression, especially with a large number of individuals in the network. I can confirm that at this scale, the coherent multiple network autoprobit model has been implemented with multiple network effects correctly recovered from simulation (see \citet{zhang2011mfaabtne} for more examples), though whether this is the proper model for potential causal mechanisms on the Framingham network is beyond the scope of this comment.

There is some mathematical consistency in the case when the network $W_{ij}$ is row-normalized, so that the total impact of one's declared friends has the same variance component for each ego. However, this implies something about the social mechanism that is not necessarily consistent with the product adoption framework. It is consistent with variants of the ``voter model'' from statistical physics \citep{Sood-Redner-voter-model}, in which one takes their preference from one of their neighbours, chosen uniformly at random, though how applicable this model is to the problem under consideration is a matter of further study.

\section{Conclusions}

The papers of CF have unambiguously demonstrated that many important behaviors are correlated on social networks, which is of considerable benefit to the measurement of nearby individuals on the network whose full traits may not have been observed. Their use of experiments to investigate small-scale contagious behaviors is equally impressive and of great benefit. It seems clear, however, that social contagion is considerably more difficult to establish, and likely more difficult than anyone thought once the FHS Social Network was processed.

Potential causative mechanisms for correlated behaviour are all around; support groups, from 12 Step Programs to The Biggest Loser TV program, have shown the world that peer pressure can be a positive force in changing unhealthy behaviour. By co-committing to a different lifestyle, or following someone else's example, people have been shown to adopt behaviours (both positive and negative). But we are still extremely far from being able to make any definitive conclusions about how these processes work without data at a much finer time scale, and social network data at much finer detail. With a large bankroll, a strong privacy policy, and a well-vetted toolkit, progress is possible.

Even when we reach this time, there are still considerable challenges to be overcome. As I have demonstrated with the QAD application to dichotomized SAR models, surprising and devastating side effects can emerge even from the most innocuous modelling choices. If theoretical derivations are not available, the least that we investigators must do is to ensure that the methods we choose do not have these unexpected properties through exhaustive simulation and the proposal of many potential alternative mechanisms -- and, most importantly, the statistical power to correctly distinguish between them.

\subsubsection*{Note On Prior Affiliation}

While I am not currently affiliated with any of the authors in any capacity, I received graduate funding from and attended weekly lab meetings with Prof. Christakis during the 2008-2009 academic year, both relating to the dichotomization of network ties \citep{thomas2010vttfl} rather than the explicit consideration of the problem of social influence. All of the work that is contained herein has taken place well after the end of this funding period.

\bibliographystyle{\baseloc statinmed}
\bibliography{\baseloc actbib}

\end{document}